\documentclass[12pt]{iopart}
\begin{document}

\title[]{Nonexistence of the non--Gaussian fixed point
predicted by the RG field theory in $4-\epsilon$ dimensions}

\author{J.~Kaupu\v{z}s}

\address{\ Institute of Mathematics and Computer Science,
University of Latvia \\ Rainja Boulevard 29, LV-1459 Riga, Latvia
\\  e--mail: \; kaupuzs@com.latnet.lv}

\begin{abstract}
The Ginzburg-Landau phase transition model is considered in
$d=4-\epsilon$ dimensions within the renormalization group (RG) approach.
The problem of existence of the non-Gaussian fixed point is discussed.
An equation is derived from the first principles of the RG theory
(under the assumption that the fixed point exists) for calculation of
the correction-to-scaling term in the asymptotic expansion of the
two-point correlation (Green's) function. It is demonstrated clearly
that, within the framework of the standard methods (well justified in
the vicinity of the fixed point) used in the perturbative RG theory,
this equation leads to an unremovable contradiction with the known RG
results. Thus, in its very basics, the RG field theory in
$4-\epsilon$ dimensions is contradictory. To avoid the contradiction,
we conclude that such a non-Gaussian fixed point, as predicted by the
RG field theory, does not exist. Our consideration does not exclude
existence of a different fixed point.
\end{abstract}




\section{Introduction}
Since the famous work of K.G.~Wilson and M.E.~Fisher~\cite{Wilson}
the renormalization group (RG) field theory has been widely
used in calculations of critical exponents~\cite{Ma,Justin}.
The basic hypothesis of this theory is the existence of a certain
fixed point for the RG transformation. However, the existence of
such a stable fixed point for the Ginzburg--Landau model (which
lies in the basis of the field theory) has not been proven
mathematically in the case of the spatial dimensionality $d<4$.
The fact that such a fixed point can be found within a scheme of
self--consistent approximations, assuming its existence
at the very beginning, cannot be regarded as a mathematical
proof. An attempt to prove the existence of the
non--Gaussian fixed point in $4-\epsilon$ dimensions has been made
in Ref.~\cite{Brydges}. The authors have considered some rather artificial
$\varphi^4$--type model, supposing that it simulates the
Ginzburg--Landau model in $4-\epsilon$ dimensions. The method of
proof is to solve the problem for a finite system (of linear
size $L$), considering the thermodynamic limit afterwards.
However, we cannot, in principle, find the non--Gaussian fixed point
at a finite $L$ for a system with real interaction, and then consider
$L \to \infty$. The problem is that, due to absence of phase transition
at $u>0$, a stable fixed point with
nonzero coupling constant $u$ of $\varphi^4$ interaction cannot exist
in a case of finite system. All such efforts to prove the existence of
the non--Gaussian fixed point, predicted in the Ginzburg--Landau model
by RG field theory, are futile, since there exists an obviously simple
proof of nonexistence presented hereafter.

\section{Fundamental equations}
We consider the Ginzburg--Landau phase transition model. The
Hamiltonian of this model in the Fourier representation reads
\begin{equation} \label{eq:H}
\frac{H}{T}= \sum\limits_{\bf k} \left( r_0+c \,{\bf k}^2 \right)
{\mid \varphi_{\bf k} \mid}^2 + uV^{-1}
\sum\limits_{{\bf k}_1,{\bf k}_2,{\bf k}_3}
\varphi_{{\bf k}_1} \varphi_{{\bf k}_2} \varphi_{{\bf k}_3}
\varphi_{-{\bf k}_1-{\bf k}_2-{\bf k}_3} \;,
\end{equation}
where $\varphi_{\bf k}=V^{-1/2} \int \varphi({\bf x})
\exp(-i {\bf kx}) \, d{\bf x}$ are Fourier components of the scalar
order parameter field $\varphi({\bf x})$, $T$ is the temperature, and
$V$ is the volume of the system. In the RG field theory~\cite{Ma,Justin}
Hamiltonian (\ref{eq:H}) is renormalized by integration of
$\exp(-H/T)$ over $\varphi_{\bf k}$ with $\Lambda/s<k<\Lambda$,
followed by a certain rescaling procedure providing a Hamiltonian
corresponding to the initial values of $V$ and $\Lambda$, where
$\Lambda$ is the upper cutoff of the
$\varphi^4$ interaction. Due to this procedure, additional terms
appear in the Hamiltonian (\ref{eq:H}), so that in general the
renormalized Hamiltonian contains a continuum of parameters.
The basic hypothesis of the RG theory in $d<4$ dimensions is the
existence of a non--Gaussian fixed point $\mu=\mu^*$ for the RG
transformation $R_s$ defined in the space of Hamiltonian parameters, i.e.,
\begin{equation} \label{eq:fixp}
 R_s \mu^* = \mu^* \;.
\end{equation}
The fixed-point values of the Hamiltonian parameters are marked by an
asterisk ($r_0^*$, $c^*$, and $u^*$, in particular). Note that
$\mu^*$ is unambiguously defined by fixing the values of $c^*$
and $\Lambda$. According to the RG
theory, the main terms in the renormalized Hamiltonian in
$d=4-\epsilon$ dimensions are those contained in (\ref{eq:H}) with
$r_0^*$ and $u^*$ of the order $\epsilon$,
whereas the additional terms are small corrections of order $\epsilon^2$.

 Consider the Fourier transform $G({\bf k}, \mu)$ of the two--point
correlation (Green's) function, corresponding to a point $\mu$.
Under the RG transformation $R_s$ this function transforms as
follows~\cite{Ma}
\begin{equation} \label{eq:RGt}
G({\bf k}, \mu)=s^{2- \eta} \, G(s {\bf k}, R_s \mu) \;.
\end{equation}
Let $G({\bf k}, \mu) \equiv G(k,\mu)$ \,
(at ${\bf k} \ne {\bf0}$ and $V \to \infty$)
be defined within $k \le \Lambda$.
Since Eq.~(\ref{eq:RGt}) holds for any
$s>1$, we can set $s= \Lambda/k$, which at $\mu = \mu^*$ yields
\begin{equation} \label{eq:asyfix}
G({\bf k},\mu^*) = a \, k^{-2 + \eta} \;\;\; for \;
k<\Lambda \;,
\end{equation}
where $a= \Lambda^{2- \eta} G(\Lambda, \mu^*)$ is the amplitude
and $\eta$ is the universal critical exponent. According to the
universality hypothesis, the infrared behavior of the Green's
function is described by the same universal value of $\eta$ at
any $\mu$ on the critical surface (with the only requirement that
all parameters of Hamiltonian (\ref{eq:H}) are present), i.e.,
\begin{equation} \label{eq:asy}
G({\bf k},\mu)= b(\mu) \, k^{-2+\eta} \;\;\; at\;\; k \to 0\;,
\end{equation}
where
\begin{equation} \label{eq:lim}
b(\mu)=\lim\limits_{k \to 0} k^{2-\eta} \, G({\bf k},\mu) \;.
\end{equation}
According to Eq.(\ref{eq:RGt}), which holds for any $s=s(k)>1$,
Eq.(\ref{eq:lim}) reduces to
\begin{equation}
b(\mu)=\lim\limits_{k \to 0} k^{2-\eta} s(k)^{2-\eta} \,
G(s {\bf k},R_s \mu) \;.
\end{equation}
By setting $s(k)=\Lambda/k$, we obtain
\begin{equation} \label{eq:b}
b(\mu)= \Lambda^{2-\eta} \, \lim\limits_{k \to 0}
G(\Lambda,R_{(\Lambda/k)} \mu)= \Lambda^{2-\eta} \,
G(\Lambda,\mu^*) \, = \, a \;,
\end{equation}
if the fixed point
$\mu^* = \lim\limits_{s \to \infty} R_s \mu$ exists.
Let us define the function $X({\bf k},\mu)$ and the self--energy
$\Sigma({\bf k},\mu)$ as follows
\begin{equation} \label{eq:X}
X({\bf k},\mu) \, = \, k^{-2} G^{-1} ({\bf k},\mu) \;,
\end{equation}
\begin{equation} \label{eq:selfe}
k^2 \, X({\bf k},\mu)= 2(r_0+ c\, {\bf k}^2) + \Sigma({\bf k},\mu) \;.
\end{equation}
Equation (\ref{eq:selfe}) is usually used in the perturbation theory,
since the self--energy has a suitable representation by Feynman diagrams.
According to Eqs.(\ref{eq:asyfix}), (\ref{eq:asy}), and
(\ref{eq:b}), we have (for $k< \Lambda$)
\begin{equation} \label{eq:Xfix}
X({\bf k},\mu^*)= \frac{1}{a} \, k^{-\eta}
\end{equation}
and
\begin{equation} \label{eq:Xcrit}
X({\bf k},\mu)= \frac{1}{a} \, k^{-\eta} \,
+ \, \delta X({\bf k},\mu) \;,
\end{equation}
where $\mu$ belongs to the critical surface,
$\mu^* = \lim\limits_{s \to \infty} R_s \mu$, and
$\delta X({\bf k},\mu)$ denotes the correction--to--scaling term.
From (\ref{eq:Xfix}) and (\ref{eq:Xcrit}) we obtain the equation
\begin{equation} \label{eq:eq}
\delta X({\bf k},\mu^* + \delta \mu) =
X({\bf k}, \mu^* + \delta \mu) - X({\bf k}, \mu^*) \; ,
\end{equation}
where $\delta \mu= \mu - \mu^*$.  This equation, of course, makes sense
only if the fixed point $\mu^*$ exists and $\mu$ includes all the
relevant Hamiltonian parameters to ensure the universal infrared
critical behavior (\ref{eq:asy}) of the correlation function.

\section{Proof of the nonexistence}
On the basis of fundamental equations obtained in the previous section,
we prove here the nonexistence of the fixed point predicted by RG
field theory, i.e., we assume the existence and derive a contradiction.
Since Eq.~(\ref{eq:eq}) is true for any small deviation $\delta \mu$
satisfying the relation
\begin{equation} \label{eq:ff}
\mu^* =\lim\limits_{s \to \infty} R_s(\mu^* +\delta\mu) \;,
\end{equation}
we choose $\delta \mu$ such that $\mu^* \Rightarrow \mu^* +
\delta\mu$ corresponds to the variation of the Hamiltonian parameters
$r_0^* \Rightarrow r_0^* + \delta r_0$,
$c^* \Rightarrow c^* + \delta c$, and
$u^* \Rightarrow u^* + \,\epsilon \times \Delta$,
where $\Delta$ is a small constant.
The values of $\delta r_0$ and $\delta c$ are choosen to fit the
critical surface and to meet the condition (\ref{eq:ff}) at fixed
$c^*=1$ and $\Lambda=1$. In particular, quantity $\delta c$ is found
\begin{equation}
\delta c \,=\, B \; \epsilon^2 \, +o(\epsilon^3) \;,
\end{equation}
with some (small) coefficient $B=B(\Delta)$, to compensate the shift in
$c$ of the order $\epsilon^2$ due to the renormalization (cf.~\cite{Ma}).
The formal $\epsilon$--expansion of $\delta X({\bf k}, \mu)$ can
be obtained in the usual way from the perturbation theory. In
this case Eq.~(\ref{eq:eq}) reduces to
\begin{equation}
\delta X({\bf k},\mu)= 2 \, \delta c + k^{-2} \,
[\delta\Sigma({\bf k},\mu) -\delta\Sigma({\bf 0},\mu)] \;,
\end{equation}
where $\delta\Sigma({\bf k},\mu)$ is the variation of self--energy
due to the substitution $\mu^* \Rightarrow \mu^* + \delta\mu$.
A simple calculation yields
\begin{equation} \label{eq:expansion}
\delta X({\bf k}, \mu)= \epsilon^2 \, [\, 2 B \,
-12 \, (2A \, \Delta + \Delta^2) \; k^{-2}
(I({\bf k}) -I({\bf 0})) \, ] \,+ o(\epsilon^3) \;,
\end{equation}
where
\begin{eqnarray} \label{eq:I}
I({\bf k}) &=& (2\pi)^{-8} \int\limits_{k_1<1} d^4 k_1
\int\limits_{k_2<1} d^4 k_2 \times k_1^{-2} k_2^{-2} \,
{\mid {\bf k}-{\bf k}_1-{\bf k}_2 \mid}^{-2} \nonumber \\
&& \times \theta(1- \mid {\bf k}-{\bf k}_1-{\bf k}_2 \mid )
\end{eqnarray}
and $A$ is the expansion coefficient in the $\epsilon$--expansion
of the renormalized coupling constant $u^*$, i.e.,
\begin{equation}
u^* = \, A \, \epsilon \, + o(\epsilon^2) \;.
\end{equation}
The theta function appears in Eq.~(\ref{eq:I}) due to the
cutting of the integration region at $k=\Lambda=1$.
Term (\ref{eq:I}) is well known~\cite{Ma}.
It behaves like \, $const + k^2 \ln k$ \, at small $k$.
The expansion coefficient at $\epsilon^2$ in Eq.~(\ref{eq:expansion})
is exact, since uncontrolled parameters of order $\epsilon^2$ contained
in the renormalized Hamiltonian $H^*$, which are absent in Eq.(\ref{eq:H}),
give a contribution of order $\epsilon^3$ to
$\delta\Sigma({\bf k},\mu)-\delta\Sigma({\bf 0},\mu)$.
In such a way, at small $k$ the expansion is unambiguous, i.e.,
\begin{equation} \label{eq:expas}
\delta X({\bf k}, \mu) = \epsilon^2 \, [\, C_1(\Delta) +
C_2(\Delta) \, \ln k \, ] \, +o(\epsilon^3) \;\;\; at \; k \to 0 \;,
\end{equation}
where $C_1(\Delta)$ and $C_2(\Delta)$ are coefficients
independent on $\epsilon$.\\
It is commonly accepted in the RG field theory to make an
expansion like (\ref{eq:expansion}), obtained from the diagrammatic
perturbation theory, to fit an asymptotic expansion at $k \to 0$,
thus determining the critical exponents.
In general, such a method is not rigorous since,
obviously, there exist such functions which do not contribute
to the asymptotic expansion in $k$ powers at $k \to 0$, but give a
contribution to the formal $\epsilon$--expansion at any fixed
$k$. Besides, the expansion coefficients do not vanish at $k \to 0$.
A trivial example of such a function is
$\epsilon^m \,[1-\tanh(\epsilon \,k^{-\epsilon})]$ where $m$ is integer.
Nevertheless, according to the general ideas of the RG theory,
in the vicinity of the fixed point the asymptotic expansion
\begin{equation} \label{eq:Xas}
X({\bf k},\mu)= \frac{1}{a} k^{-\eta} +
b_1 k^{\epsilon+o(\epsilon^2)} + b_2 k^{2+o(\epsilon)} + ...
\end{equation}
is valid not only at $k \to 0$, but within $k<\Lambda$. The latter means
that terms of the kind $\epsilon^m \,[1-\tanh(\epsilon \,k^{-\epsilon})]$
are absent or negligible.
In such a way, if there exists a fixed point, then we can
obtain correct $\epsilon$--expansion of $\delta X({\bf k},\mu)$
at small $k$ by expanding the term $b_1 k^{\epsilon+o(\epsilon^2)}$
(with $b_1=b_1(\epsilon,\Delta)$) in
Eq.~(\ref{eq:Xas}) in $\epsilon$ powers, and the result must
agree with (\ref{eq:expas}) at small $\Delta$, at least.
The latter, however, is impossible
since Eq.~(\ref{eq:expas}) never agree with
\begin{equation}
\delta X({\bf k},\mu) = b_1(\epsilon,\Delta) \, [\, 1+ \epsilon \ln k
+o(\epsilon^2) \, ]
\end{equation}
obtained from (\ref{eq:Xas}) at $k \to 0$. Thus, we have
arrived to an obvious contradiction, which means that the initial
assumption about existence of a certain fixed point, predicted
by the RG field theory in $4-\epsilon$ dimensions, is not valid.
The only reason why this contradiction has not been detected
before, seems, is the fact that Eq.(\ref{eq:eq}) never has been
considered in literature. Since the results of the RG field theory in
$4- \epsilon$ dimensions completely are based on the formal
$\epsilon$--expansion, the predicted "fixed point", obviously,
is a set of Hamiltonian parameters at which Eq.~(\ref{eq:fixp}) is
satisfied in the limit $\epsilon \to 0$ at a fixed $s$,
but is not satisfied in the limit $s \to \infty$ at a fixed $\epsilon$.

\section{Conclusions}
We have demonstrated clearly that, in its very basics, the RG field
theory in $4-\epsilon$ dimensions is contradictory.
Based on a mathematically correct (according to the general arguments
of the RG theory) method, we have shown that such a
fixed point in $4-\epsilon$ dimensions, as predicted by the RG
field theory, does not exist. It should be noticed, however, that our
consideration does not exclude existence of a different fixed point.

\References
\bibitem[1]{Wilson} K.G.~Wilson, M.E.~Fisher, {\it
Phys.Rev.Lett.} {\bf 28}, 240 (1972)
\bibitem[2]{Ma} Shang--Keng Ma, {\it Modern Theory of Critical
Phenomena} (W.A.~Benjamin, Inc., New York, 1976)
\bibitem[3]{Justin} J.~Zinn--Justin, {\it Quantum Field Theory and
Critical Phenomena} (Clarendon Press, Oxford, 1996)
\bibitem[4]{Brydges} D.~Brydges, J.~Dimock, T.R.~Hurd, {\it archived as
96-681 of $mp_{-}arc$ in Texas}
\endrefs

\end{document}